\documentclass[12pt]{article}
\pdfoutput=1
\usepackage{epsfig}\parskip 5pt plus 1pt
\usepackage{bbm}
\usepackage{array}
\usepackage{hhline}
\usepackage{amsmath}
\usepackage{amssymb}
\usepackage{amsfonts}
\usepackage{mathrsfs}
\usepackage{graphicx}
\usepackage{multicol}
\usepackage{fancybox}
\usepackage{pifont}
\usepackage{lscape}
\usepackage{color}
\usepackage{ulem}
\usepackage{cite}

\allowdisplaybreaks[1]

\newcommand{\be}{\begin{equation}}
\newcommand{\ee}{\end{equation}}
\newcommand{\bea}{\begin{eqnarray}}
\newcommand{\eea}{\end{eqnarray}}

\newcolumntype{M}[1]{>{\centering}m{#1}}

\newcommand{\equ}[1]{\eq~(\ref{equ:#1})}
\newcommand{\figu}[1]{\fig~\ref{fig:#1}}

\newcommand{\ie}{{\it i.e.}}

\newcommand{\eg}{{\it e.g.}}

\newcommand{\cf}{{\it cf.}}

\newcommand{\eq}{Eq.}

\newcommand{\fig}{Fig.}

\newcommand{\Ref}{Ref.}
\newcommand{\Refs}{Refs.}

\newcommand{\Tab}{Table}

\begin{document}
\vspace*{-1cm}
%\phantom{hep-ph/yymmnnn}
%\hfill{LOCAL-PREPRINT-NUMBER}
{\begin{flushright}
KA-TP-30-2012\\
MPP-2012-114\\
SFB/CPP-12-50
\end{flushright}}
\vskip 2.0cm
\begin{center}
{\Large\bf Interpretation of precision tests in the Higgs sector in terms of physics beyond the Standard Model}
\end{center}
\vskip 1.0  cm
\begin{center}
{\large Florian Bonnet}$\,^a$\footnote{bonnet@pd.infn.it}, \,{\large Toshihiko Ota}$\,^b$\footnote{toshi@mppmu.mpg.de},\,
{\large Michael Rauch}$\,^c$\footnote{rauch@particle.uni-karlsruhe.de},\,
{\large Walter Winter}$\,^a$\footnote{winter@physik.uni-wuerzburg.de}\\
\vskip .5cm
$^a\,$Institut f\"{u}r Theoretische Physik und Astrophysik, Universit\"{a}t W\"{u}rzburg, Germany \\
\vskip .2cm
$^b\,$Max-Planck-Institut f\"{u}r Physik (Werner-Heisenberg-Institut),  M\"{u}nchen, Germany\\
\vskip .2cm
$^c\,$Institut f\"{u}r Theoretische Physik, Karlsruhe Institute of Technology (KIT), Germany\\
\end{center}
\vskip 0.5cm

\begin{center}
\end{center}

\begin{abstract}

%The discovery of the Higgs boson opens the possibility of precision physics of the Higgs sector. 
We demonstrate how the measurements of the Higgs-fermion and Higgs-gauge boson couplings can be interpreted in terms of physics beyond the Standard Model in a model-independent way. That is, we describe deviations from the Standard Model by effective $d=6$ operators made of Higgs fields and gauge fields, under the hypothesis that the new physics may show up in the Higgs sector only and the effective operators are generated at tree level. While the effective operator coefficients are independent in general, the completion of the theory at high energies will lead to specific correlations which will be recovered between Higgs-fermion and Higgs-gauge boson couplings. We demonstrate that the current measurement of these couplings in terms of tree-level new physics requires several new mediators with specific relationships among different couplings.
New insights in the effective theory and mediator spaces can be expected for improved measurements from the inclusive $H \rightarrow \tau \tau$ and the exclusive vector boson fusion-dominated $H \rightarrow \gamma \gamma$ search channels, as well as the measurement of the Higgs self-couplings, including higher order couplings which do not exist in the Standard Model.
\end{abstract}

\newpage
\setcounter{footnote}{0}

\section{Introduction} 

%(Higgs discovery, literature for measurement of couplings [see Ellis paper],
%composite Higgs and relationship to that,  our old paper and what is new here
%etc.)

The last update of the ATLAS and CMS collaborations on the Higgs searches
established the existence of a new resonance at 125 GeV with $5\sigma$
significance for the signal~\cite{ATLASJuly4,CMSJuly4}. This resonance 
is compatible with the Higgs boson, that would complete the Standard Model (SM) picture.
However, to be sure that this new resonance is the SM Higgs boson, one need,
among other things, measure its couplings and compare them to their SM
prediction.  In case of inconsistencies, one should try to identify the
information that can be drawn on new physics from the Higgs data.
Several studies have been performed on Monte-Carlo
expectations~\cite{Zeppenfeld:2000td,Duehrssen:CSC,Duhrssen:2004cv,Lafaye:2009vr,Bock:2010nz,Peskin:2012we}
as well as on the first hints from 2011
data~\cite{Giardino:2012ww,Carmi:2012yp,Espinosa:2012ir,Ellis:2012rx,Azatov:2012bz,Azatov:2012rd,Farina:2012ea,Klute:2012pu,Espinosa:2012vu,Carmi:2012zd,Bertolini:2012gu}.
In the light of the new Higgs data
several new studies have already appeared where general Higgs couplings are
probed~\cite{Corbett:2012dm,Giardino:2012dp,Carmi:2012,Low:2012rj,Montull:2012ik,Espinosa:2012,Banerjee:2012xc}.
These studies use a model independent approach by adding a small, independent,
deviation to each of the Higgs couplings to the SM particles. 

However, one
needs to interpret these deviations in terms of possible new physics models.
With that goal, we study in this paper the implication of a set of effective
operators on the Higgs data, see \Refs~\cite{Hagiwara:1993qt,Hagiwara:1996kf,Eboli:1998vg,GonzalezGarcia:1999fq,Barger:2003rs,Dutta:2008bh,Kanemura:2008ub,delAguila:2010mx,Bonnet:2011yx} for earlier works, and interpret them with respect to their possible ultraviolet completions. As it
was shown in  \Ref~\cite{Bonnet:2011yx}, correlations between the coefficient of
the effective operators arise when considering these new physics
models, which offer the possibility to test these models via the measurement of
the Higgs couplings. The paper is organized as follows. We first review the
formalism developed in \Ref~\cite{Bonnet:2011yx}, showing how the Higgs
couplings and observables are modified by the presence of effective operators.
We then apply the recent constraints obtained at the LHC to the coefficient of
the effective operators and show how they can be used to discriminate between
different new physics model. Finally we propose new ways to improve the
measurement of the effective coefficients considered in this work.

\section{Modification of the Higgs sector from effective operators} 

Effective operators are useful to describe Beyond the SM (BSM) physics in the low-energy limit in a model-independent way, similar to Fermi's theory for beta decay. The effective Lagrangian can be written as
\begin{equation}
\mathcal{L}_{\mathrm{eff}}=\mathcal{L}_{\mathrm{SM}}+\sum_i \alpha_i \mathcal{O}_i\,,
\label{eq:Leff}
\end{equation}
where $\alpha_i \propto \Lambda^{d-4}$ is suppressed by the scale of new
physics $\Lambda$ and the mass dimension $d$ of the operator. While the lowest dimensional operator, the unique $d=5$ Weinberg operator, violates lepton number and leads to Majorana neutrino masses, there is a phletora of $d=6$ operators which can be constructed from SM fields~\cite{Buchmuller:1985jz} (see also \Ref~\cite{Grzadkowski:2010es} for the discussion of the number of independent operators). 
Among these operators, only a few can be built from the Higgs and SM gauge fields only, which can be directly related to observables in the Higgs sector. Following \Ref~\cite{Bonnet:2011yx}, we focus on the ones which can be mediated at tree level (see \Refs~\cite{Arzt:1993gz,Arzt:1994gp}), assuming that these may be the leading ones in perturbative theories.\footnote{For a discussion how well a different set of the Higgs-gauge interactions can be measured, see \Ref~\cite{Corbett:2012dm}.}
In the Buchm{\"u}ller-Wyler basis~\cite{Buchmuller:1985jz},  they read
\begin{eqnarray}
\mathcal{O}_{\phi}=-\frac{1}{3}(\phi^{\dagger}\phi)^3 \,,&& \boldsymbol{\mathcal{O}_{\partial\phi}=\frac{1}{2}\partial_{\mu}(\phi^{\dagger}\phi)\partial^{\mu}(\phi^{\dagger}\phi)}\,,\label{equ:ophidphi}\\
\boldsymbol{\mathcal{O}_{\phi}^{(1)}=(\phi^{\dagger}\phi)(D_{\mu}\phi)^{\dagger}(D^{\mu}\phi)}\,, &&\mathcal{O}_{\phi}^{(3)}=(\phi^{\dagger}D_{\mu}\phi)((D^{\mu}\phi)^{\dagger}\phi)\, , \label{equ:ophi1ophi3}
\end{eqnarray}
where $\phi$ is the Higgs doublet.
As demonstrated in \Ref~\cite{Bonnet:2011yx}, $\mathcal{O}_{\phi}$ can only be tested by measuring the Higgs self-couplings~\cite{Djouadi:1999rca,Baur:2002rb,Baur:2002qd,Baur:2003gp,Dolan:2012rv,Boudjema:1995cb,Djouadi:1999gv,Baur:2009uw} (or new interactions, such as higher order interactions of the Higgs). Since this may be difficult at the LHC, we disregard it for the moment and comment on it later. In addition,  $\mathcal{O}_{\phi}^{(3)}$ is severely constrained by electroweak precision tests (EWPT), in particular, the contribution to the T~parameter, which leads to $\alpha_\phi^{(3)} v^2 \lesssim 3 \cdot 10^{-4}$. Therefore, any completion of the theory must not violate this bound, which restricts the possible mediators leading to the effective operators $\mathcal{O}_{\partial\phi}$ and $\mathcal{O}_{\phi}^{(1)}$~\cite{Bonnet:2011yx}.

In order to describe the modification of the SM Lagrangian by
$\mathcal{O}_{\partial\phi}$ and $\mathcal{O}_{\phi}^{(1)}$, all SM
relationships are expressed in terms of the best measured quantities,
the Fermi constant $G_F$, the fine-structure constant $\alpha$, the mass
of the $Z$ boson $M_Z$, and the mass of the Higgs boson $M_H$, which is
accessible at the LHC, and renormalizing the Lagrangian by keeping these fixed
to their measured values. This procedure is in detail discussed in
\Ref~\cite{Bonnet:2011yx}, and leads (among other implications) to the
following  leading order modifications of the SM Higgs-gauge boson and
Higgs-fermion couplings:
\begin{eqnarray}
\lambda_{HVV}&=&\lambda_{HVV_{\mathrm{SM}}} \left( 1+\alpha_{\phi}^{(1)}\frac{v^2}{2}-\alpha_{\partial\phi}\frac{v^2}{2} \right)\,,\label{equ:HVV}\\
\lambda_{Hff} & = &
\frac{Y_{f}}{\sqrt{2}}
=
\frac{Y_{f_{\mathrm{SM}}}}{\sqrt{2}} \left( 1-
\alpha_{\partial\phi}\frac{v^2}{2} \right)\,.
\label{equ:fermion}
\end{eqnarray}
Here $V$ represents the SM gauge bosons $W$ and $Z$. Because of these modifications, the couplings of the Higgs boson to two gluons, mediated by a fermion loop, is also modified and goes like
\begin{equation}
\lambda_{H{\rm gg}} 
=
\lambda_{H{\rm gg}_{\rm{SM}}} \left(1-
\alpha_{\partial\phi}\frac{v^2}{2} \right)\,,
\end{equation}
while the coupling to two photons reads\footnote{Here we only consider the tree
level correction, and we assume that the loop contribution induced by the same
mediators is sub-leading. }
\begin{equation}
\lambda_{H\gamma\gamma}=\lambda_{H\gamma\gamma_{\rm{SM}}}\frac{(1-\alpha_{\partial\phi}\frac{v^2}{2})\frac{4}{3}A_{1/2}^H(\tau_t)+(1+\alpha_{\phi}^{(1)}\frac{v^2}{2}-\alpha_{\partial\phi}\frac{v^2}{2})A_{1}^H(\tau_W)}{\frac{4}{3}A_{1/2}^H(\tau_t)+A_{1}^H(\tau_W)}\,,
\label{equ:Hgammagamma}
\end{equation}
where $\tau_i=\frac{M_H^2}{4M_i^2}$, $A^H_{1/2}(\tau)=2[\tau+(\tau-1)f(\tau)]\tau^{-2}$, $A^H_1(\tau)=-[2\tau^2+3\tau+3(2\tau-1)f(\tau)]\tau^{-2}$ and
\begin{eqnarray}
f(\tau)=\left\{ \begin{array}{cc} \rm{arcsin}^2\sqrt{\tau} & \tau\leq1 \\
-\frac{1}{4}\left[\log\left(\frac{1+\sqrt{1-\tau^{-1}}}{1-\sqrt{1-\tau^{-1}}}\right)-i\pi\right] & \tau>1\end{array}\right.\,.
\end{eqnarray}
For the comparison with the literature, we also use the deviations from the SM couplings  $c_V \equiv \lambda_{HVV}/\lambda_{HVV_{\mathrm{SM}}}$ and $c_F \equiv Y_{f}/Y_{f_{\mathrm{SM}}}$ in the following.
As a consequence, the effective operator coefficients can be easily rewritten as
\begin{eqnarray}
  \alpha_{\phi}^{(1)}\frac{v^2}{2} & = & c_V-c_F  \, ,\label{equ:alphaphi1} \\
\alpha_{\partial\phi}\frac{v^2}{2} & = &  1 - c_F \, \, \label{equ:alphapartialphi}
\end{eqnarray}
in terms of the deviations of the gauge boson and fermion couplings. Note that in the above formulas we have already assumed that $\alpha_\phi^{(3)} \simeq 0$ .

\begin{table}[t!]
\centering
\begin{tabular}{|l|p{2.8cm}||c|c|c||c|c||c|c|}
\hline
 & & & & & \multicolumn{2}{c||}{Gauge} & \multicolumn{2}{c|}{Non-gauge} \\
Coeff. & Participating in & ${\bf 1}^s_0$ & ${\bf 3}^s_0$ & ${\bf 3}^s_1$ & ${\bf 1}^v_0$  & ${\bf 3}^v_0$  & ${\bf \tilde 1}^v_0$  & ${\bf \tilde 3}^v_0$\\
\hline
$\alpha_{\phi}^{(1)} $ & $HWW$, $HZZ$ & 0 & $2\frac{\mu_{\Delta}^2}{m_{\Delta}^4}$ & $4\frac{|\mu_{\Delta_1}|^{2}}{m_{\Delta_1}^4}$ & 
0 & $- \frac{g_U^2}{2 m_U^2}$ & 0 & $-2 \frac{\lambda_{U}^2}{m_U^2}$ \\
\hline
$\alpha_{\phi}^{(3)}$ & EWPT! & 0 & $-2\frac{\mu_{\Delta}^2}{m_{\Delta}^4}$ & $4\frac{|\mu_{\Delta_1}|^{2}}{m_{\Delta_1}^4}$ & $-2 \frac{g_V^2}{m_V^2}$ & 0 & 0 & $2 \frac{\lambda_{U}^2}{m_U^2}$ \\
\hline
$\alpha_{\partial\phi} $ & $HWW$, $HZZ$, $H \bar f f$ & $\frac{\mu_S^2}{m_S^4}$ & $\frac{\mu_{\Delta}^2}{ m_{\Delta}^4}$ & $0$ & $\frac{g_V^2}{m_V^2}$ & $\frac{g_U^2}{4 m_U^2}$ & $- \frac{\lambda_{V}^2}{m_V^2}$ & $- \frac{\lambda_{U}^2}{m_U^2}$\\
\hline
\end{tabular}
\caption{\label{tab:combin} List of mediators and their contributions to the effective operators 
$\mathcal{O}_{\phi}^{(1)}$, $\mathcal{O}_{\phi}^{(3)}$, and $\mathcal{O}_{\partial\phi}$,  (only single mediator cases). The operator $\mathcal{O}_{\phi}^{(3)}$ is limited by electroweak precision tests (EWPT). Table from \Ref~\cite{Bonnet:2011yx}.
}
\end{table}

\section{Interpretation of deviations from the SM}

The question that naturally arises is the interpretation of the
effective operators $\mathcal{O}_{\partial\phi}$ and
$\mathcal{O}_{\phi}^{(1)}$ in terms of new physics, and if correlations
among these operators can appear in specific theories.  Focusing on
SU(2) singlet, doublet, and triplet scalars, and singlet and triplet
vectors as the mediation fields of the effective operators, all possibilities
for the tree-level mediation of these operators have been identified in
\Ref~\cite{Bonnet:2011yx}. 
The procedure can be qualitatively described as follows: all possible
topologies and fields insertions, which can lead to
the effective operators, have been identified, together with the minimum 
set of necessary interactions to generate the operators (plus interactions
among the mediators). Then the
fields have been simultaneously integrated out to derive the operator
coefficients and possible effects of new mediator-mediator couplings
in the multi-mediator cases. 
From this minimum set of interactions, no four-fermion
vertices can be generated because scalar/vector-fermion couplings
have not been introduced.

In order to simplify the presentation, we have assigned symbols to the
mediators and list the SM quantum numbers in brackets in the form ${\bf
X}^{\mathcal{L}}_Y$, where
\begin{itemize}
\item
 ${\bf X}$ denotes the SU(2) nature, i.e., singlet ${\bf 1}$, doublet ${\bf 2}$, or triplet ${\bf 3}$.
\item
     $\mathcal{L}$ refers to the Lorentz nature, i.e., scalar ($s$) 
     and vector ($v$).
\item
 $Y$ refers to the hypercharge $Y=Q-I^W_3$.
\end{itemize}
The result is summarized in \Tab~\ref{tab:combin} for the single mediator cases, \ie, when only one mediator is present at a time. For multiple new mediators, the interactions add trivially. The different couplings in the table refer to couplings between the mediators and the Higgs, and are in detail specified in \Ref~\cite{Bonnet:2011yx}. As a peculiarity in the table, the new vector mediators may be gauge fields, or the Lagrangian results from a broken gauge symmetry (non-gauge vectors), where the interactions are introduced in a different way. Furthermore note that the operator $\mathcal{O}_{\phi}$ in \equ{ophidphi} may be mediated by a doublet scalar, which does not appear in \Tab~\ref{tab:combin} because it does not lead to any of the shown operators.

%${\bf 1}^s_0$ & ${\bf 3}^s_0$ & ${\bf 3}^s_1$ & ${\bf 1}^v_0$  & ${\bf 3}^v_0$  & ${\bf \tilde 1}^v_0$  & ${\bf \tilde 3}^v_0$

From \Tab~\ref{tab:combin}, one can easily read off that 
\begin{itemize}
\item
 The single mediator cases with ${\bf 3}^s_0$, ${\bf 3}^s_1$, ${\bf 1}^v_0$, and ${\bf \tilde 3}^v_0$ are forbidden, since they produce large corrections to the electroweak precision data via $\alpha_{\phi}^{(3)}$.
\item
 If cancellations among the contributions from more than one mediator are allowed, the contribution to $\alpha^{(3)}_\phi$ can be eliminated by choosing the couplings accordingly. The most prominent case frequently used in the literature \cite{Georgi:1985nv,Chanowitz:1985ug,Gunion:1989ci,Gunion:1990dt,Logan:2010en} and, in fact, the only one we study here, is that of two triplet scalars ${\bf 3}^s_0$ and ${\bf 3}^s_1$. The cancellation condition $-2 \mu_{\Delta}^2/m_{\Delta}^4 + 4 |\mu_{\Delta_1}|^{2}/m_{\Delta_1}^4=0$ implies that $\alpha_{\phi}^{(1)}=4 \mu_\Delta^2/m_\Delta^4$ and  $\alpha_{\partial \phi}=\mu_\Delta^2/m_\Delta^4$.
\item 
 For each mediator, particular correlations between the effective
      operator coefficients  $\alpha_{\phi}^{(1)}$ and $\alpha_{\partial
      \phi}$ are obtained. For example, for ${\bf 3}^v_0$, one has
      $\alpha_{\phi}^{(1)}= -2 \alpha_{\partial \phi}$. These
      correlations will be directly translated into correlations between $c_V$ and $c_F$, which are not present if only the effective field theory is considered. In addition, the sign of the deviation from these couplings contains meaningful information. 
\end{itemize}

\begin{figure}[t]
\begin{center}
\includegraphics[width=\textwidth]{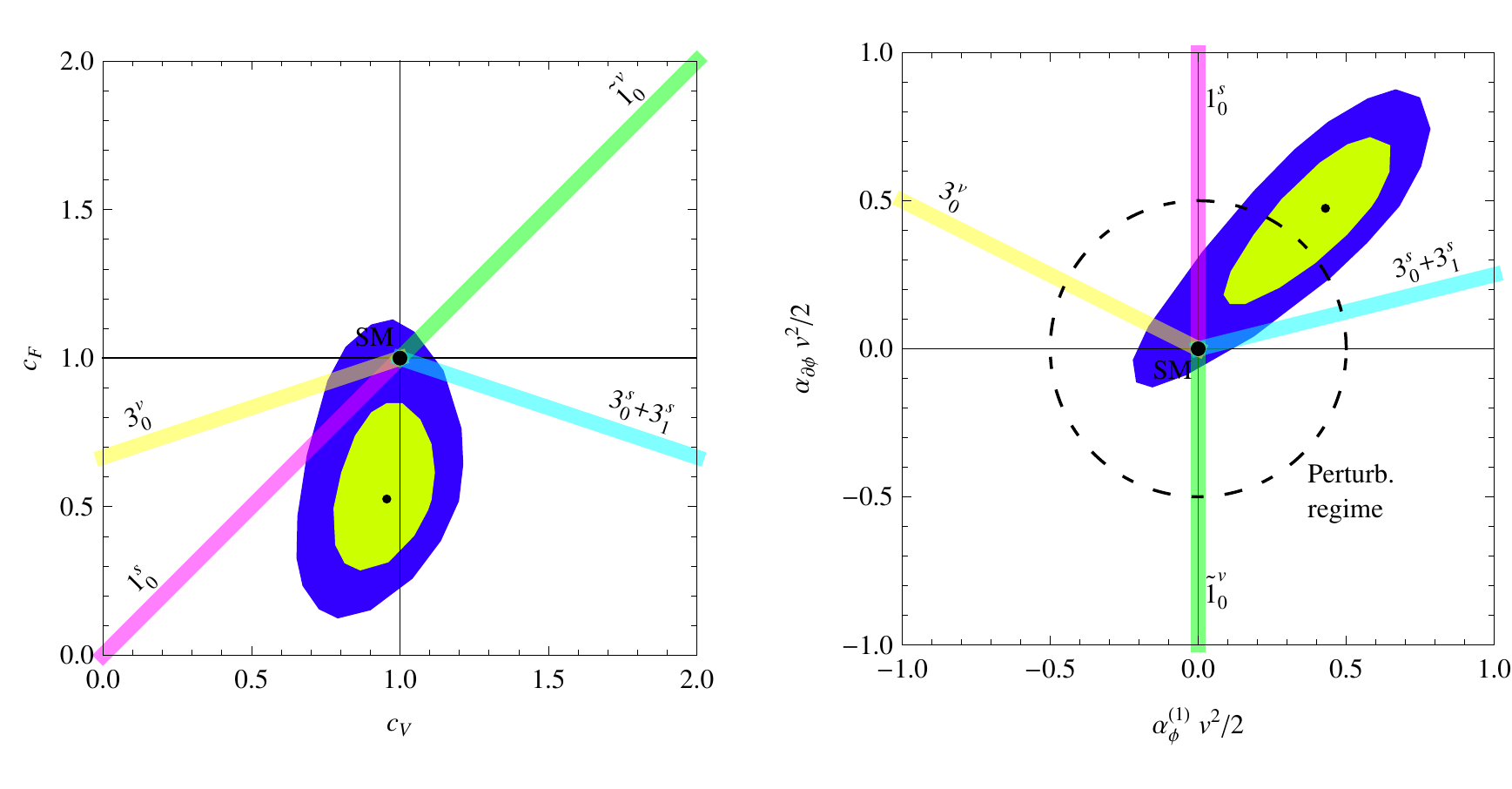}
\end{center}
\caption{\label{fig:couplings} Filled contours: measurement of the Higgs-vector
boson and Higgs-fermion couplings in terms of $c_V$ and $c_F$ (left panel,
``experimental plane''), and $\alpha_\phi^{(1)}$ and  $\alpha_{\partial \phi}$
(right panel, ``theory plane''). The data are at the 68\% and 95\% confidence
levels, taken from \Ref~\cite{CMSJuly4}. The different lines (rays) represent
the correlations among the effective operator coefficients for specific
mediators, where in the case of small couplings the SM is recovered. The dashed
circle in the right panel shows the perturbative regime (geometric average of
couplings smaller than about 0.5), \ie, the region where the effective operator
approach is expected to be a good description.
} 
\end{figure}

We show in \figu{couplings}, left panel, the actual (first) measurement of
$c_V$ and $c_F$ by the CMS collaboration and the translation into the effective
operator plane, right panel, using \equ{alphaphi1} and
\equ{alphapartialphi}.\footnote{For recent combined analyses of the Higgs
self-couplings, including the contribution of different channels and
constraints from ATLAS, see \eg\
\Refs~\cite{Carmi:2012,Ellis:2012,Gillioz:2012se}.} The different lines (rays)
represent the correlations among the effective operator coefficients for
specific mediators, where in the case of small couplings ($\alpha \rightarrow
0$ or, equivalently, $c \rightarrow 1$), the SM is recovered. The dashed circle
in the right panel shows the perturbative regime, \ie, the region where the
effective operator approach is expected to be a good description. Note that the
case ${\bf 1}^s_0$ corresponds to the minimal composite Higgs model MCHM4
\cite{Agashe:2004rs,Espinosa:2010vn} in the positive coupling constant region,
since they both produce only $\alpha_{\partial\phi}$.

From this figure, one can read off that while the SM is disfavored at the 68\%
CL, a simple interpretation with the current data in terms of the simplest
possible models seems unlikely. However, using, for instance, three scalars,
one can construct more complicated models aligned with the fit contour (\cf,
right panel). Such a model would also require specific relationships among the
couplings, to give an idea for the required complexity.  However, future
measurements will improve the precision, and the best-fit shape may change. The
picture in \figu{couplings} can be used to directly interpret these
measurements in terms of theory. The left panel in \figu{couplings} can be
regarded as ``experimental plane'', and the right panel as ``theory plane''. In
addition note, as we will show below, that for $m_H \simeq 125 \, \mathrm{GeV}$
the information from the leading channels is highly correlated in the effective
operator (theory) plane (\cf, Fig.~5 in \Ref~\cite{Bonnet:2011yx}). This
correlation can be easily seen in the right panel of \figu{couplings}.

Note that in the CMS analysis, the parameters $c_V$ and $c_F$ are independent,
and therefore a genuine two parameter fit is performed. In terms of the
mediators at tree level, the EWPT constraint turns this fit into a one
parameter problem even for two mediators, and the fit of the couplings is
reduced to a fit along the lines in \figu{couplings} -- which, however, depends
on the mediator. The 95\% confidence level constraints on the individual
coupling-mass ratios in \Tab~\ref{tab:combin} can be directly read off from
\figu{couplings} (ranges from zero to the point where the lines cross the
blue/dark region, to be corrected for one versus two d.o.f. - which leads to a
somewhat smaller range than shown in the figure). 

\section{Perspectives for different search channels}

The LHC data are collected through different search channels. The ratio of the
event rate compared to its SM prediction, $R_X^Y$, of a  channel where the
Higgs is produced by the mechanism $X$ and where it decays into a given final
state $Y$ is:
\begin{equation}
R_X^Y=\frac{\sigma(X\rightarrow H)\times BR(H\rightarrow Y)}{\sigma_{\mathrm{SM}}(X\rightarrow H)\times BR_{\mathrm{SM}}(H\rightarrow Y)}\,.
\end{equation}
Currently the production mechanisms considered are gluon fusion ($X$=gg),
vector boson fusion ($X$=VBF) and Higgsstrahlung/associated production
($X$=VH), while the decays to gauge bosons ($WW$, $ZZ$, $\gamma\gamma$) and
fermions ($\tau\tau$, $bb$) are measured.  In the most recent data, the
channels studied are: Inclusive (``incl'') channels, which are dominated by the
gluon fusion mechanism even for large values of $\alpha_{\partial\phi}$ and
$\alpha_{\phi}^{(1)}$. In the numerical analysis below we add the
sub-leading contributions from the other production modes blindly,
i.e.~assuming that acceptance and efficencies are identical for the
different production modes.
The other possibility are exclusive channels, which only rely on one
production mode. For VBF a non-negligible part stems from gluon-fusion
production and this is added using the fractions given in the
experimental analyses~\cite{CMSVBFgamma,CMSVBFtau}.
The selection of the best search channels does not only depend
on the Higgs mass, but also on other factors, such as energy resolution and SM
backgrounds. The leading search channels right now are the $H \rightarrow
\gamma \gamma$ and $H \rightarrow ZZ$ channels, in spite of these branching
ratios not being the dominant ones, for which the observables can be
approximated (as inclusive channels) by $R_{\mathrm{incl}}^{\gamma \gamma}$ and
$R_{\mathrm{incl}}^{ZZ}$. 

\begin{figure}[t]
\begin{center}
\includegraphics[width=10cm]{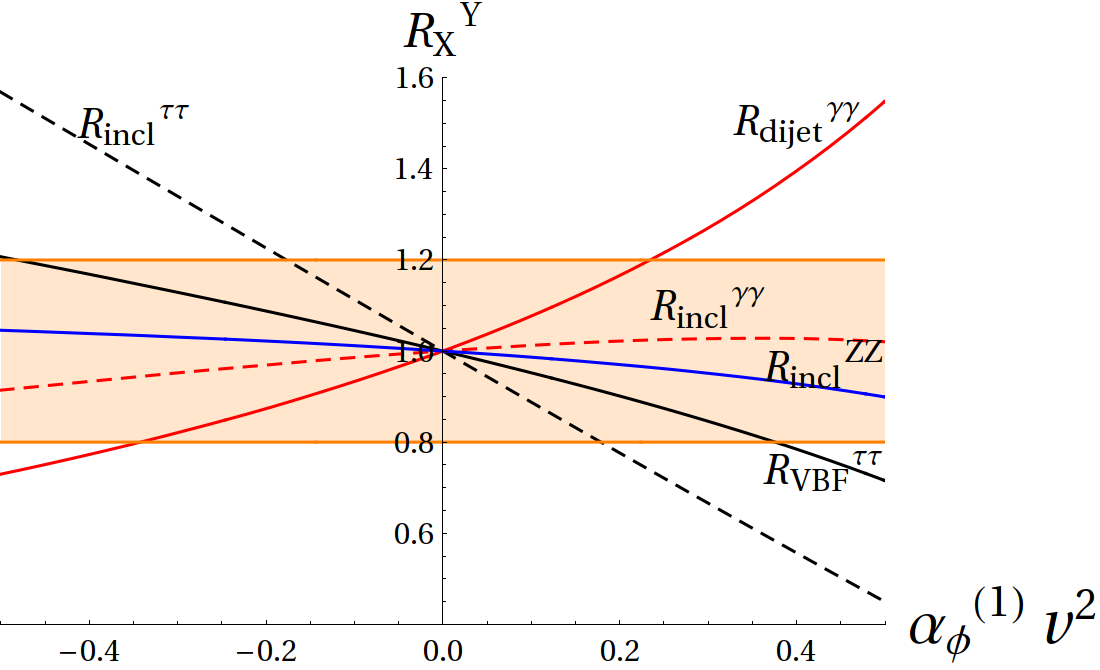}
\end{center}
\caption{\label{fig:R} Value of the individual effective operator coefficient
$\alpha_{\phi}^{(1)}\frac{v^2}{2}$ as a function of $R_X^Y$ for different
search channels for a fixed relationship
$\alpha_{\partial\phi}=\alpha_{\phi}^{(1)}$. The orange shading illustrates the
effect of a 20\% measurement of $R_X^Y$.}
\end{figure}

From \figu{couplings}, left panel, one can read off that $c_V$ is better
measured than $c_F$ at the moment, which translates through \equ{HVV} to the
degeneracy $\alpha_\phi^{(1)} \propto \alpha_{\partial \phi}$ in the right
panel. This means that the individual effective operator coefficients are
difficult to extract from this combination.  So what observables can be used to
break this degeneracy?  In order to address this question, we show in \figu{R}
the value of  $R_X^Y$  as a function of the individual effective operator
coefficient $\alpha_{\phi}^{(1)}v^2$, using a fixed relationship
$\alpha_{\partial\phi}=\alpha_{\phi}^{(1)}$ (\ie, along the correlation axis in
\figu{couplings}, right panel). As one can read off from \figu{R},
$R_{\mathrm{incl}}^{\gamma \gamma}$ and $R_{\mathrm{incl}}^{ZZ}$ indeed exhibit
very little dependence on the individual coefficient $\alpha_{\phi}^{(1)}$,
which means that a very small error on $R$ is needed to constrain this
coefficient. This is illustrated by the gray-shaded area, which shows the
effect of a 20\% measurement of $R_X^Y$: in this case, the individual
coefficient cannot be constrained.

Consider, for instance, $R_{\mathrm{incl}}^{ZZ}$. One can read off from
\equ{HVV} and \equ{fermion} that
\begin{equation}
 R_{\mathrm{gg}}^{ZZ} \propto \left( 1 - \alpha_{\partial \phi} v^2 \right) \left( 1 + \alpha_\phi^{(1)} v^2 - \alpha_{\partial \phi}  v^2 \right)\times\frac{\Gamma_{\rm{tot}_{SM}}}{\Gamma_{\rm{tot}}} \, ,
\end{equation}
where the total decay width contains also a dependence in
$\alpha_{\partial\phi}$ and $\alpha_{\phi}^{(1)}$ \footnote{One can make the
approximation: $\frac{\Gamma_{\rm{tot}}}{\Gamma_{\rm{tot}_{SM}}}\simeq
1+0.25(\alpha_{\phi}^{(1)}v^2-\alpha_{\partial\phi}v^2)-0.75\alpha_{\partial\phi}v^2$}.
This means that  it only measures a combination of  $\alpha_\phi^{(1)}$
and $\alpha_{\partial \phi}$. 
In this channel the rate change in gluon-fusion production is mostly
compensated by a corresponding change in the total width, originating
from the dominating fermionic decay modes, while the combination of
$\alpha_\phi^{(1)}$ and $\alpha_{\partial \phi}$ leaves the decay width
unchanged.
From \equ{HVV} and \equ{fermion}, it is obvious that
$R_{\mathrm{VBF}}^{\tau \tau}$ measures the same combination of
parameters, where the role of the new couplings between production and
decay is exchanged.  Therefore, its role in \figu{R} as a useful
observable is also limited, in addition to the low statistics expected
in this exclusive channel. 

The most efficient observable is the gluon fusion dominated ratio
$R_{\mathrm{incl}}^{\tau \tau}$, since it measures $c_F$, and therefore
$\alpha_{\partial \phi}$, exclusively; \cf, \equ{fermion}. It clearly exhibits
the strongest  dependence on $\alpha_\phi^{(1)} =\alpha_{\partial \phi}$ in
\figu{R}. Another option to break the degeneracy is to use a vector boson
fusion dominated exclusive $R_{\mathrm{VBF}}^{\gamma \gamma}$ channel: we show
$R_{\mathrm{dijet}}^{\gamma \gamma}$ (dominated by VBF) in \figu{R}. 
Here the different dominant Higgs production mode (compared to
$R_{\mathrm{incl}}^{\gamma \gamma}$) leads to a different dependence on the
effective operator coefficients. It is, at the end, a matter of statistics
where the best information can be obtained. From the Higgs production cross
sections at $125 \, \mathrm{GeV}$, which is suppressed for vector boson fusion
compared to gluon fusion by about one order of magnitude, these options should
be comparable in terms of statistics, but it is not clear how this comparison
looks like in the presence of  selection efficiencies, backgrounds, and energy
resolution. In principle, $R_{\mathrm{incl}}^{\tau \tau}$ is the channel from
which the strongest constraints are expected, since the dependence in \figu{R}
is stronger, which means that it is worth to optimize it. In terms of theory,
both channels will lead to strong constraints of the couplings in
\Tab~\ref{tab:combin}, and may finally rule out the SM prediction.

Apart from the mediators listed in \Tab~\ref{tab:combin}, a doublet scalar $\boldsymbol{2}^s_{1/2}$ has been found as a possible mediator of $\mathcal{O}_{\phi}$ in \equ{ophidphi}~\cite{Bonnet:2011yx}, which does not necessarily have to take a vev. The operator $\mathcal{O}_{\phi}$ will lead to shifts of the Higgs self-couplings, which are known to be hard to measure at the LHC, for instance
\begin{equation}
\lambda_{HHH} = \lambda_{HHH_{\mathrm{SM}}}(1-\alpha_{\partial\phi}\frac{v^2}{2}+\frac{1}{3}\alpha_{\phi}\frac{v^2}{\lambda_{0}})\, ,
\label{equ:HHH}
\end{equation}
where $\lambda_0=M_H^2/v^2$. If $\alpha_{\partial\phi}$ can be measured from the Higgs-fermion couplings, then $\alpha_{\phi}$ can be in principle extracted from measuring this coupling.
However, note that also new higher order effective Higgs interactions are generated by  $\mathcal{O}_{\phi}$: 
\begin{equation}
\delta \mathcal{L} =
-\alpha_{\phi}\frac{v}{4} H^5-\frac{\alpha_{\phi}}{24} H^6\, .
\end{equation}
These are not present in the SM and may be useful in the future to discover new
physics  and interpret it in terms of BSM theories. However, there will be also
an irreducible SM background from the production of several Higgses by SM
interactions, and it remains to be seen if useful information can be obtained
from these couplings. In addition, the measurement of these couplings requires
beyond LHC technology, such as a linear or muon collider.

\section{Summary and conclusions}

The discovery of the Higgs at the LHC is the beginning of a new era which can be regarded as the precision physics of the Higgs sector. Especially the new data allow to probe the values of the Higgs-gauge boson and Higgs-fermion couplings. We have focused on the theoretical interpretation of possible deviations from the Standard Model identified there, where we have used a model-independent approach. We have tested the hypothesis that new physics may lead to effective $d=6$ operators made from Higgs and gauge bosons only, and that the new mediators lead to a tree level generation of such an operator. Two of the operators can be directly related to a linear combination of the Higgs-gauge boson and Higgs-fermion couplings, which are perhaps the simplest observables in the Higgs sector. 

We have demonstrated how current bounds translate into the measurement of the effective operator coefficients.
With the data from the leading channels, an approximate degeneracy between the coefficients has been identified. Future measurements of sub-leading channels, such as  the inclusive $H \rightarrow \tau \tau$ channel and the exclusive vector boson fusion-dominated $H \rightarrow \gamma \gamma$ channel, 
can be used to reduce that and to exclude the SM case.  We have also demonstrated how the measurements can be interpreted in terms of the tree level mediators, which lead to particular correlations in the effective operator plane. Although the actual low statistics does not allow one to draw definitive conclusions, we have shown that the LHC is powerful enough to constrain the effective operators that modify the Higgs sector.
Current data do not allow for a simple interpretation of the possible deviation from the Standard Model in terms of a single mediator. We have illustrated that a beyond the Standard Model interpretation of the current deviation requires several new fields with particular relationships among the couplings, which does not seem appealing. 

Finally, additional information on possible new physics in the Higgs sector can
be obtained from measuring the Higgs self-couplings, which is known to be
challenging. As an alternative, higher order effective Higgs couplings can be
used to measure the remaining effective operator, which may be indicative for a
new scalar doublet coupling to the Higgs sector only. Although these may be
more difficult to measure, this possibility should be studied in the future.

\subsubsection*{Acknowledgments}

We would like to thank M. B. Gavela for useful and illuminating discussions.
FB and WW would like to acknowledge support from DFG grants WI 2639/3-1 and WI
2639/4-1. We would also like to thank the Galileo Galilei Institute for Theoretical Physics for the hospitality during our stay, where a substantial fraction of this work was carried out.
MR acknowledges support by the Deutsche Forschungsgemeinschaft via the
Sonderforschungsbereich/Transregio SFB/TR-9 ``Computational Particle
Physics'' and the Initiative and Networking Fund of the Helmholtz
Association, contract HA-101(``Physics at the Terascale'').

%%%%%%%%%%%%%%%%%%
%                                                           %
%                 bibliography                   %
%                                                           %
%%%%%%%%%%%%%%%%%%

\bibliographystyle{h-physrev5}
%\bibliography{references}

\end{document}